\documentclass[conference]{IEEEtran}
\IEEEoverridecommandlockouts
\usepackage{subcaption}
\usepackage{colortbl}
\usepackage{booktabs}
\usepackage{siunitx}
\usepackage{multirow}
\usepackage{cite}
\usepackage{amsmath,amssymb,amsfonts}
\usepackage{algorithmic}
\usepackage{graphicx}
\usepackage{textcomp}
\usepackage{xcolor}
\usepackage{booktabs}

\usepackage{caption} 
\def\BibTeX{{\rm B\kern-.05em{\sc i\kern-.025em b}\kern-.08em
    T\kern-.1667em\lower.7ex\hbox{E}\kern-.125emX}}
\begin{document}

\title{FADEL: Uncertainty-aware Fake Audio Detection with Evidential Deep Learning\\
\thanks{This work was supported by the COMPA grant funded by the Korea government (MSIT and Police) (No. RS2023-00235082).}
}

\author{
    \IEEEauthorblockN{Ju Yeon Kang$^{\star}$ \qquad Ji Won Yoon$^{\dagger}$ \qquad Semin Kim$^{\star}$ \qquad Min Hyun Han$^{\star}$ \qquad Nam Soo Kim$^{\star}$}
    \IEEEauthorblockA{
        $^{\star}$ Department of Electrical and Computer Engineering and INMC, Seoul National University, Seoul, South Korea \\
        $^{\dagger}$ Department of AI, Chung-Ang University, Seoul, South Korea \\
        Email: \{jykang, smkim21, mhhan\}@hi.snu.ac.kr, jiwonyoon@cau.ac.kr, nkim@snu.ac.kr
    }
}

\maketitle

\begin{abstract}

Recently, fake audio detection has gained significant attention, as advancements in speech synthesis and voice conversion have increased the vulnerability of automatic speaker verification~(ASV) systems to spoofing attacks.
A key challenge in this task is generalizing models to detect unseen, out-of-distribution~(OOD) attacks.
Although existing approaches have shown promising results, they inherently suffer from overconfidence issues due to the usage of softmax for classification, which can produce unreliable predictions when encountering unpredictable spoofing attempts.
To deal with this limitation, we propose a novel framework called fake audio detection with evidential learning~(FADEL).
By modeling class probabilities with a Dirichlet distribution, FADEL incorporates model uncertainty into its predictions, thereby leading to more robust performance in OOD scenarios.
Experimental results on the ASVspoof2019 Logical Access~(LA) and ASVspoof2021 LA datasets indicate that the proposed method significantly improves the performance of baseline models.
Furthermore, we demonstrate the validity of uncertainty estimation by analyzing a strong correlation between average uncertainty and equal error rate~(EER) across different spoofing algorithms.

\end{abstract}


\section{Introduction}
Recent advancements in deep learning have shown remarkable performance in speech synthesis, allowing systems to generate speech that closely mimics human voices\cite{tacotron,fastspeech, lorenzo2018voice, zhao2020voice}.
However, the progress has increased the vulnerability of automatic speaker verification~(ASV) systems to advanced spoofing attacks\cite{advancesantispoof, vulnerability}.
To ensure the reliability of ASV systems, it is essential to develop fake audio detection~(a.k.a audio anti-spoofing) systems that effectively distinguish between genuine and spoofed speech.

A key challenge in fake audio detection is that the anti-spoofing countermeasures~(CMs) must confront a wide range of unseen attacks that were not encountered during training \cite{paul2017generalization,zhang2021empirical,cohen2022study}. 
As sophisticated spoofing techniques continue to emerge, CMs should effectively generalize to these evolving threats. Therefore, anti-spoof datasets, such as ASVspoof2019\cite{2019asvspoof} and ASVspoof2021\cite{2021asvspoof}, incorporate out-of-distribution~(OOD) spoofing scenarios, where the training and test sets include different types of attacks. The approach evaluates the models' ability to detect unpredictable attacks, providing a critical measure of how well the models can generalize to new, real-world spoofing attempts.

Recent anti-spoofing approaches have made significant advances in addressing the challenges posed by OOD spoofing attacks\cite{yi2023audio, li2024audio}. 
A range of models have been developed, leveraging architectures such as CNN\cite{dinkel2017end, chintha2020recurrent,hua2021towards}, RawNet2\cite{tak2021end, wang2023rawnet}, GNN\cite{rawgat, aasist}, ResNet\cite{hua2021towards}, DARTS\cite{ge2021raw}, and Transformer\cite{liu2023leveraging}, along with various training strategies\cite{chen2020generalization,zhang2021one,ma2021continual,tak2022rawboost,shim2023multi}.
However, existing methods have limitations in dealing with the overconfidence problem inherent in the softmax function, especially in the context of OOD spoofing attacks\cite{nguyen2015deep, guo2017calibration}.
A common practice in state-of-the-art CMs is to use the softmax function to convert the model's final outputs into class probabilities, followed by optimization using maximum likelihood estimation~(MLE), such as cross entropy loss.
Since the softmax function employs an exponential operation to convert logits into probabilities, small differences in logits can be exaggerated, leading to overly confident predictions for certain classes.
Moreover, MLE does not consider the uncertainty of the prediction, as it focuses solely on maximizing the probability of the target class.
As a result, softmax can produce high-confidence predictions even when the model is uncertain or incorrect, especially in OOD scenarios.
Given the importance of accurately detecting OOD spoofing attacks in fake audio detection CMs, addressing the overconfidence problem is critical for ensuring system reliability.


To mitigate this overconfidence issue, one possible solution is evidential deep learning~(EDL)\cite{sensoy2018evidential}. 
EDL aims to quantify the model's uncertainty in predictions by modeling the distribution of class probabilities using a Dirichlet distribution.
The model's output is interpreted as ``evidence'' that reflects the confidence level in each class.
The evidence serves as a parameter for the Dirichlet distribution, which in turn models uncertainty.
During the training stage, EDL learns to generate evidence from the input data, enabling the model to estimate the uncertainty and incorporate it into the prediction process.


\begin{figure*}[htbp]
\begin{center}
\includegraphics[width=\textwidth]{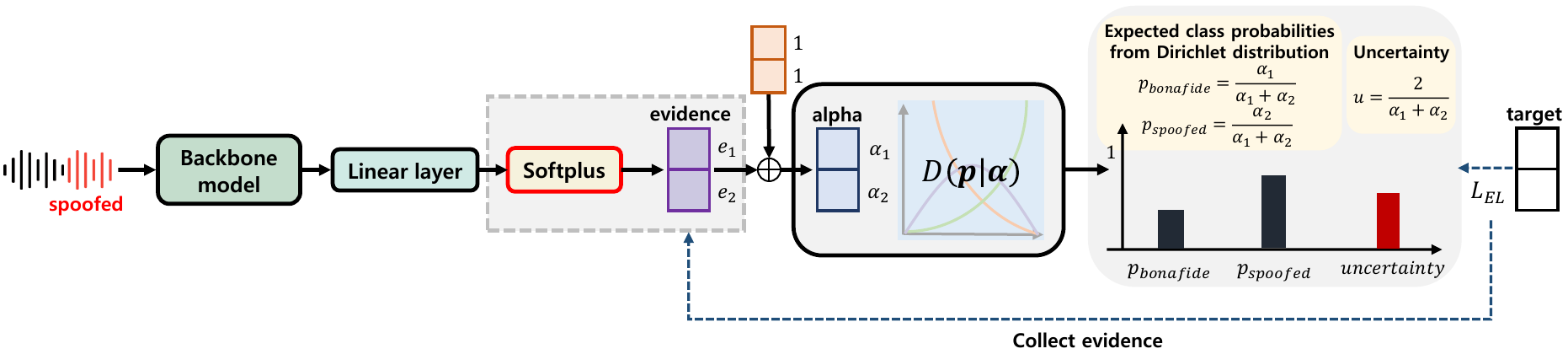}
\end{center}
\caption{An Overview of FADEL.}
\label{overall}
\end{figure*}

In this paper, we propose a novel framework called fake audio detection with evidential learning, namely FADEL, which can effectively improve detection performance by addressing the limitations of softmax in OOD scenarios.
Specifically, FADEL derives class probabilities from the mean of the Dirichlet distribution, inherently accounting for the model’s uncertainty in its predictions.
This integration of model uncertainty mitigates the softmax overconfidence and enables more robust predictions when facing OOD spoofing attacks.
Experimental results show that FADEL significantly improves the performance of fake audio detection baselines on the ASVspoof2019 Logical Access~(LA) dataset. 
In cross-dataset evaluation, where the model is trained on ASVspoof2019 LA and tested on ASVspoof2021 LA, FADEL also outperforms other approaches.
Additionally, we demonstrate the validity of the uncertainty by showing a strong alignment between the quantified average uncertainty and the equal error rate~(EER) for each spoofing algorithm.
\textit{To the best of our knowledge, this is the first attempt to incorporate model uncertainty into fake audio detection systems with evidential learning.}


\section{Proposed Method}
In this section, we describe our FADEL framework, an uncertainty-aware training scheme for fake audio detection.
An overview of FADEL is illustrated in Fig. \ref{overall}.

\subsection{Motivation}\label{Motivation}
Existing state-of-the-art fake audio detection CMs use the softmax function to transform the model outputs into class probabilities. 
For the $i$-th pair $(\mathbf{x_i}, \mathbf{y_i})$ in the training dataset, where $\mathbf{x_i}$ is the input utterance and $\mathbf{y_i}$ is one hot vector for the target label, the probability $\mathbf{p_i}$ is calculated as follows:

\begin{equation}
p_{ij} = \frac{\exp(z_{ij})}{\sum_{k=1}^{K} \exp(z_{ik})},
\end{equation}

for each class $j$, where $\mathbf{z_i}$ is the model's output logit, and $K$ is the number of classes. 
The probability $\mathbf{p_i}$ is then used to compute the objective function.
In typical anti-spoofing datasets, such as ASVspoof2019\cite{2019asvspoof} and ASVspoof2021\cite{2021asvspoof}, the number of spoofed examples significantly outweighs the number of genuine examples. 
To address this class imbalance, CMs employ a weighted cross-entropy loss~(WCE). 
The WCE loss is defined as:

\begin{equation}
L_{WCE} = -\sum_{j=1}^{K} w_j \cdot y_{ij} \log(p_{ij}),
\end{equation}
where $w_j$ is the weight assigned to each class. 

Despite its effectiveness, the training scheme with softmax function and WCE can lead to an overconfidence problem in predictions.
The exponential nature of the softmax function can amplify small differences in logits, assigning excessively high probabilities to certain classes.
WCE loss only maximizes the target class probability without considering the uncertainty of predictions.
Therefore, in OOD spoofing attack scenarios where the model is uncertain about its predictions, the issue can be problematic, potentially misleading predictions.

\subsection{Evidential Deep Learning Framework}\label{2-B}
EDL\cite{sensoy2018evidential} aims to estimate uncertainty in predictions by modeling the Dirichlet distribution.
The Dirichlet distribution is a distribution over class probabilities, which is suitable for capturing uncertainty.
The specific framework is as follows.

For a given input $\mathbf{x_i}$ and \(K\) classes, the model output is treated as evidence vector  \(\mathbf{e_i} = [e_{i1}, e_{i2}, \dots, e_{iK}]\), which represents the model's confidence in each class.
Since evidence is a non-negative value, activation functions such as ReLU, softplus, or exponential function are applied to the model output.
This evidence $\mathbf{e_i}$ is then utilized to define the parameters of the Dirichlet distribution, $\boldsymbol{\alpha_i}$, as:
\begin{equation}
\alpha_{ik} = e_{ik} + 1,
\end{equation}
for each class $k$.
Finally, the resulting Dirichlet distribution \(\mathbf{p_i} \sim \text{Dir}(\boldsymbol{\alpha_i})\) where $\mathbf{p_i}$ is class probabilities, is capable of modeling the uncertainty in its predictions.

Based on the theory of Subjective Logic\cite{jsang2018subjective}, the uncertainty \(u_i\) for input $\mathbf{x_i}$ is defined as:
\begin{equation}
u_i = \frac{K}{\sum_{k=1}^{K} \alpha_{ik}},
\end{equation}
where $\sum_{k=1}^{K} \alpha_{ik}$ is the total sum of the Dirichlet parameters across all classes.
The uncertainty \(u_i\) is inversely proportional to the total evidence, reflecting how confident the model is about its prediction.

\subsection{Class Probabilities with Uncertainty}\label{2-C}
FADEL leverages the EDL framework to address the overconfidence problem inherent in softmax by incorporating the model's uncertainty into its predictions. 
To capture the uncertainty in predictions, FADEL defines its class probabilities using the mean of the Dirichlet distribution. 
For a given sample $\mathbf{x_i}$, the mean of the Dirichlet distribution for each class \(k\) is calculated as:
\begin{equation}
\bar{p_{ik}} = \frac{\alpha_{ik}}{\sum_{j=1}^{K} \alpha_{ij}},
\label{eq5}
\end{equation}
where \(\alpha_{ik}\) represents the Dirichlet parameter for class \(k\).
Eq. \eqref{eq5} ensures that the class probabilities $\bar{\mathbf{p_i}}$ reflect the inherent uncertainty in the predictions.

\subsection{Training Objective}
The objective function of FADEL follows the principles of EDL, which considers the Dirichlet distribution as a prior.
Our objective function for the sample $\mathbf{x_i}$ is defined as:
\begin{align}
\mathcal{L}_{EL} &= \int \left[ \sum_{j=1}^{K} -w_j \cdot y_{ij} \log(p_{ij}) \right] \frac{1}{B(\boldsymbol{\alpha_i})} \prod_{j=1}^{K} p_{ij}^{\alpha_{ij}-1} \text{d}\mathbf{p_i} \\
&= \sum_{j=1}^{K} w_j \cdot y_{ij} \left( \psi(S_i) - \psi(\alpha_{ij}) \right),
\end{align}
where $B$ is the beta function, $\psi$ is the digamma function, and $S_i = \sum_{j=1}^{K} \alpha_{ij}$.
By learning the objective function with respect to $\boldsymbol{\alpha_i}$, the model is trained to generate high evidence for the correct class while minimizing evidence for incorrect classes.

\section{Experimental setting}

\subsection{Dataset and Metrics}\label{AA}

We conducted the experiments using the train, development, and evaluation sets of the ASVspoof2019 LA\cite{2019asvspoof} dataset.
To assess the model's performance against OOD spoofing attacks, the evaluation set includes spoofing attacks not seen during training. 
Specifically, the training and development sets contain six spoofing algorithms (A01-A06), while the evaluation set introduces 13 additional algorithms (A07-A19).
In cross-dataset evaluation, we employed the ASVspoof2019 LA train and development sets for training and ASVspoof2021 LA\cite{2021asvspoof} evaluation set for testing.

For evaluation, we used Equal Error Rate~(EER) and minimum normalized Tandem Detection Cost Function~(t-DCF)\cite{tdcf}.
t-DCF evaluates the combined performance of the ASV system and the CM, considering the impact of the CM on the ASV system's reliability.

\subsection{Implementation Details}
We applied FADEL to two backbone models, Res-TSSDNet\cite{hua2021towards} and AASIST\cite{aasist}, utilizing their original architecture and hyperparameter configurations.
For the activation functions of the FADEL framework, we experimented with exponential, ReLU, and softplus functions. 

We compared Res-TSSDNet-FADEL and AASIST-FADEL with previous competing approaches, including PC-DARTS\cite{ge2021raw}, Res-TSSDNet, RawNet2\cite{tak2021end}, RawGAT-ST\cite{rawgat}, AASIST.
To ensure a fair comparison, the baseline models were trained using weighted cross-entropy loss. 
The weight ratio between the spoof and bonafide classes was set at 1:9.
For the cross-dataset evaluation in Table \ref{cross}, we employed ASAM\cite{shim2023multi} as a baseline, a recently proposed method that has demonstrated strong performance in improving the generalization ability of backbone models.
We applied ASAM to AASIST and trained it solely on the ASVspoof2019 LA dataset, without using a multi-dataset.

All experiments were conducted over 100 epochs using a single NVIDIA GeForce RTX 2080 Ti GPU and a single NVIDIA Titan RTX GPU. 
Since previous researches\cite{wang2021comparative, kanervisto2021optimizing} have shown that performance can vary significantly depending on the random seed, we conducted all experiments three times with different seeds and reported both the average and best results.


\begin{figure*}[htbp]
\begin{subfigure}{0.32\linewidth}
\centering
\includegraphics[width=\textwidth]{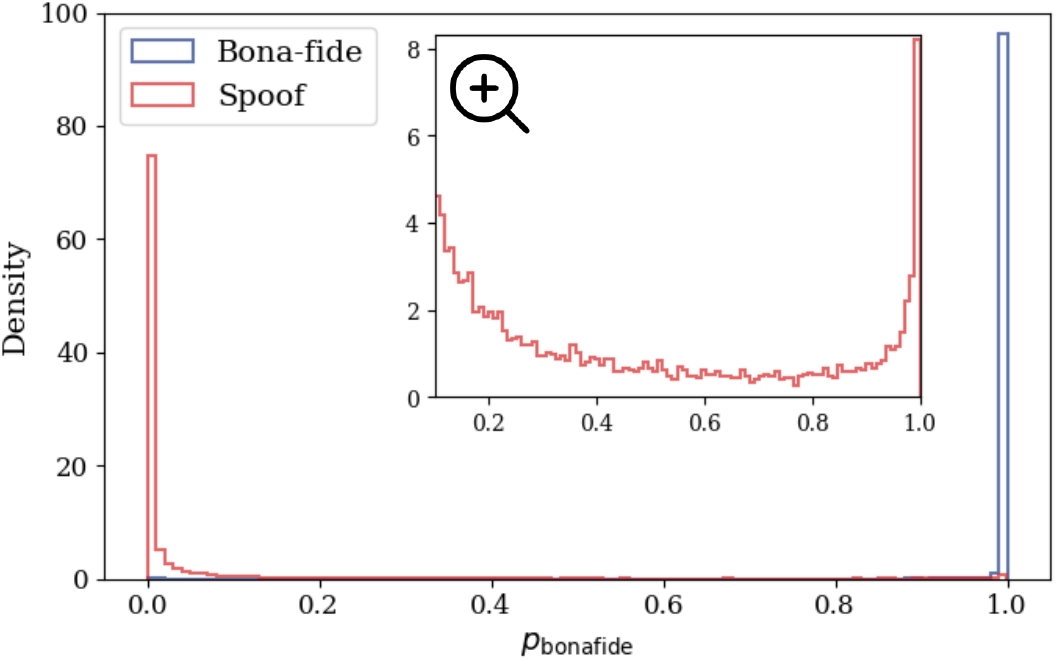}
\caption{AASIST.}
\label{fig2a}
\end{subfigure}
\begin{subfigure}{0.32\linewidth}
\centering
\includegraphics[width=\textwidth]{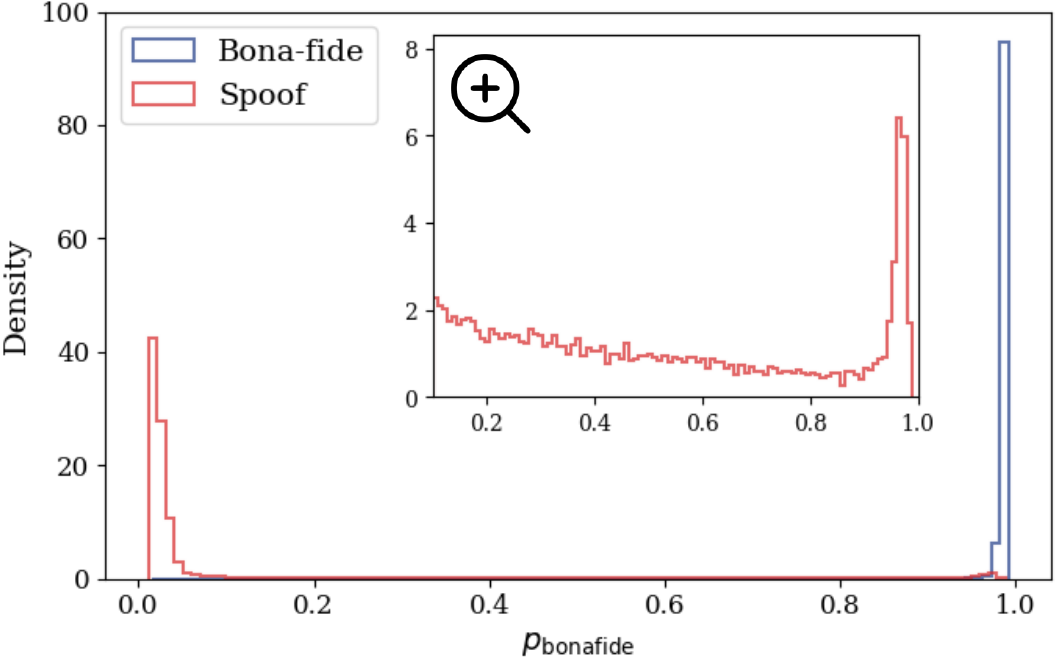}
\caption{AASIST-FADEL.}
\label{fig2b}
\end{subfigure}
\begin{subfigure}{0.32\linewidth}
\centering
\includegraphics[width=\textwidth]{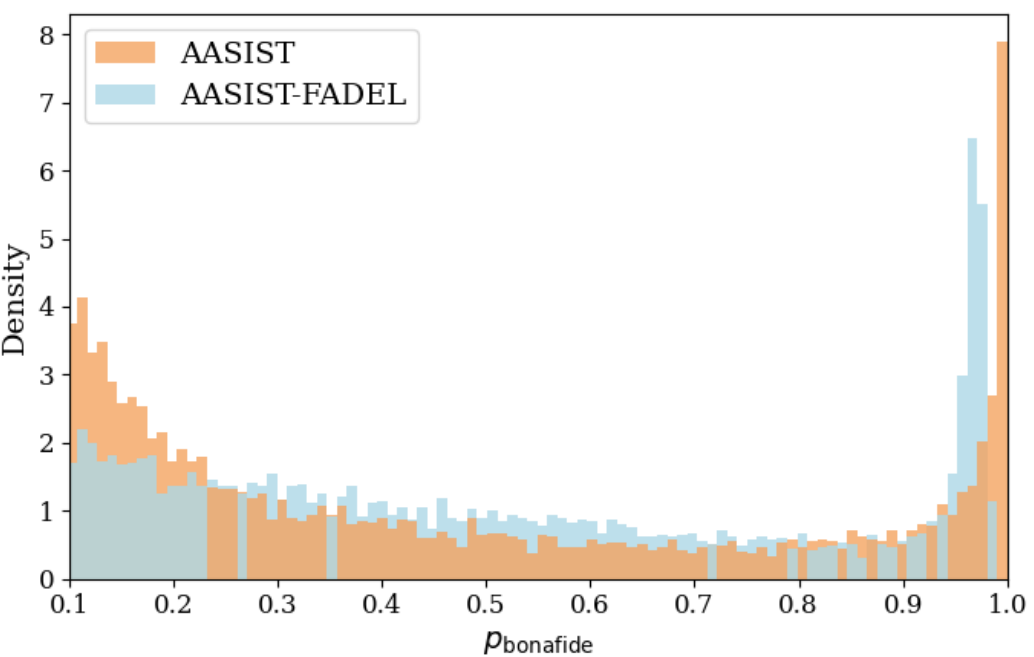}
\caption{Comparison of spoof predictions.}
\label{fig2c}
\end{subfigure}
\caption{Histograms of probabilities for the bonafide class, $p_{bonafide}$, predicted by AASIST and AASIST-FADEL: 
Insets in (a) and (b) present a detailed view of spoof predictions between 0.1 and 1.
(c) provides a comparison of spoof predictions between AASIST and AASIST-FADEL in a probability range of 0.1 to 1.}
\label{over}
\end{figure*}


\begin{table}[]
\caption{Results for the ASVspoof2019 LA evaluation set.}
\begin{center}
\begin{tabular}{lcccc}
\toprule
                                  & \multicolumn{2}{c}{\textbf{EER(\%)}} & \multicolumn{2}{c}{\textbf{min t-DCF}} \\ \cmidrule{2-5}
\multirow{-2}{*}{\textbf{System}} & avg               & best             & avg                & best              \\ \midrule
LFCC+PC-DARTS\cite{ge2021raw}                     & 5.57              & 5.43             & 0.1021             & 0.0885            \\ \cmidrule{1-5}
CQT+2D-Res-TSSDNet\cite{hua2021towards}                & 5.38              & 4.72             & 0.1347             & 0.1268            \\ \cmidrule{1-5}
RawNet2\cite{tak2021end}                           & 3.97              & 3.55             & 0.1088             & 0.0892            \\ \cmidrule{1-5}
RawGAT-ST\cite{rawgat}                         & 1.61              & 1.41             & 0.0498             & 0.0458            \\ \cmidrule{1-5}
Res-TSSDNet\cite{hua2021towards}                       & 3.53              & 3.01             & 0.1093             & 0.0932            \\
\rowcolor[gray]{0.9}
\textbf{+ Ours, FADEL}            & 2.92              & 2.79             & 0.0878              & 0.0875            \\ \cmidrule{1-5}
AASIST\cite{aasist}                            & 1.47              & 1.35             & 0.0464             & 0.0375            \\
\rowcolor[gray]{0.9}
\textbf{+ Ours, FADEL}            & \textbf{1.21}     & \textbf{1.18}    & \textbf{0.0340}    & \textbf{0.0276}   \\ \bottomrule
\end{tabular}
\end{center}
\label{main}
\end{table}


\begin{table}[]
\caption{Results for the ASVspoof2021 LA evaluation set, with systems trained on the ASVspoof2019 LA train and development set.}
\begin{center}
\begin{tabular}{lcccc}
\toprule
                                  & \multicolumn{2}{c}{\textbf{EER(\%)}} & \multicolumn{2}{c}{\textbf{min t-DCF}} \\ \cmidrule{2-5}
\multirow{-2}{*}{\textbf{System}} & avg               & best             & avg                & best              \\ \midrule
AASIST\cite{aasist}                     & 8.08              & 7.36             & 0.4037             & 0.3764            \\ \cmidrule{1-5}
+ ASAM\cite{shim2023multi}                     & 6.10              & 5.16             & 0.3448             & 0.3244            \\
\rowcolor[gray]{0.9}
\textbf{+ Ours, FADEL}            & \textbf{5.60}     & \textbf{4.91}    & \textbf{0.3334}    & \textbf{0.3108}   \\ \bottomrule
\end{tabular}
\end{center}
\label{cross}
\end{table}

\section{Result and Analysis}
\subsection{Main Results}
The results in Table \ref{main} clearly demonstrate that integrating FADEL with both Res-TSSDNet and AASIST leads to significant improvements in performance.
For Res-TSSDNet with FADEL, we observed a 17\%  reduction in EER and a 20\% reduction in min t-DCF compared to the baseline Res-TSSDNet. 
Similarly, for AASIST with FADEL, EER improved by 18\% compared to the baseline, while min t-DCF showed a 27\% improvement.


To further evaluate the model's generalization ability, Table \ref{cross} presents the cross-dataset evaluation results. 
AASIST-FADEL demonstrated significant improvements over AASIST in both EER and min t-DCF, indicating that incorporating model uncertainty leads to more robust performance in OOD spoofing scenarios.
Furthermore, AASIST-FADEL achieved better performance than the ASAM method.


\subsection{Overconfidence Problem}
Fig. \ref{fig2a} and Fig. \ref{fig2b} present histograms of the probabilities for the bonafide class, $p_{bonafide}$, predicted by the AASIST and AASIST-FADEL, respectively. 
In the AASIST histogram, samples are heavily concentrated in the edge bins~(near 0 and 1).
In contrast, the edge bins in AASIST-FADEL are much less populated, with probabilities distributed more evenly compared to AASIST.
The comparison of spoof predictions in Fig \ref{fig2c} further 
emphasizes overly confident predictions of AASIST.

The difference arises from the maximum likelihood estimation~(MLE) of AASIST, which encourages binary predictions and tends to assign probabilities near 0 or 1.
However, AASIST-FADEL is designed not only for binary classifications but also for capturing model uncertainty in the predictions.
The model generates predictions that reflect the varying degrees of the model's uncertainty about each sample, thereby avoiding extreme probabilities when the model is uncertain.

Notably, in Fig. \ref{fig2b}, while some spoof samples are assigned relatively high $p_{bonafide}$, they are generally lower than those of actual bonafide samples.
This demonstrates that AASIST-FADEL can distinguish between bonafide and realistic-looking spoof samples by reflecting different confidence levels.
As a result, AASIST-FADEL reduces misclassifications, especially in challenging cases, leading to better performance in EER and min t-DCF.

\begin{figure}[htbp]
\centerline{\includegraphics[width=0.5\textwidth]{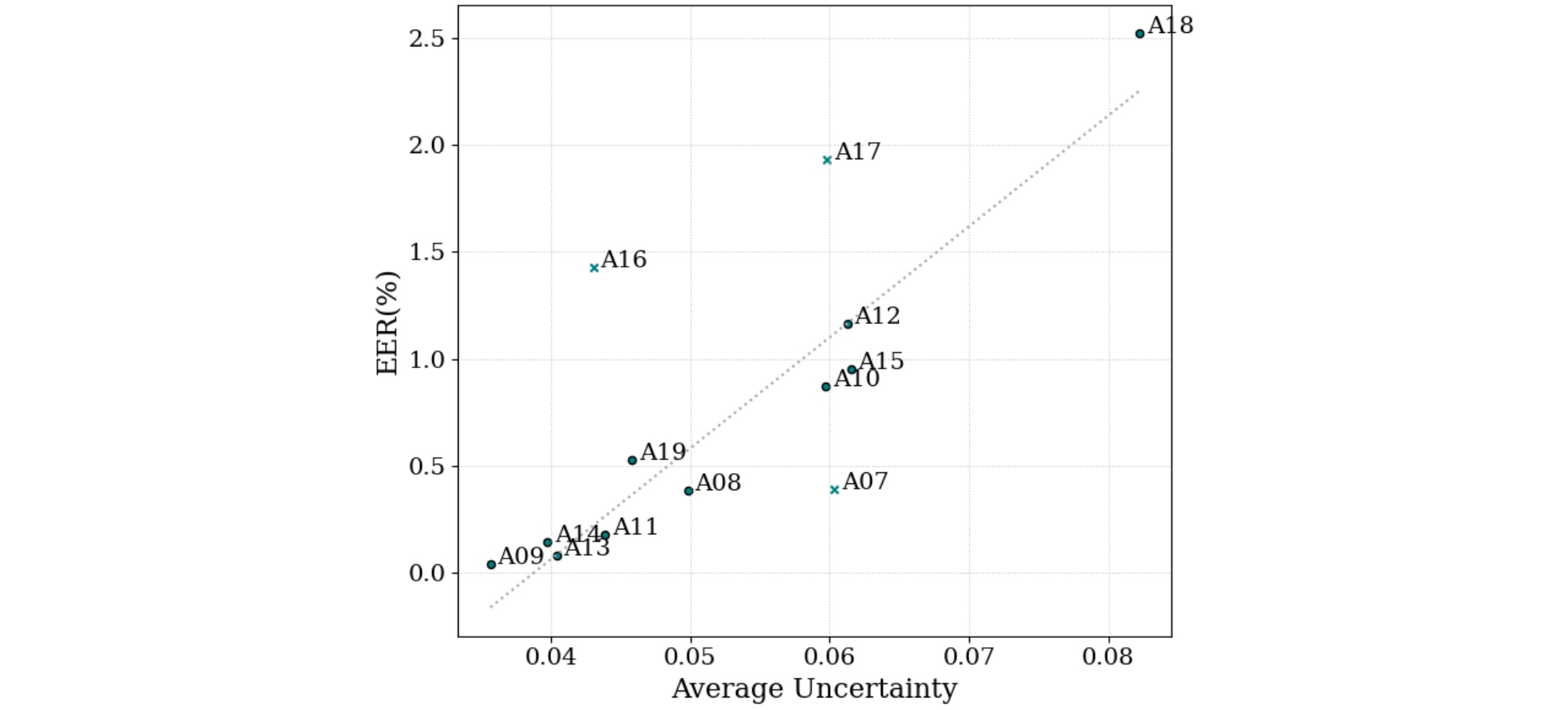}}
\caption{Scatter plot showing the relationship between average uncertainty and EER across spoofing algorithms in the ASVspoof2019 LA evaluation set. 
Algorithms with lower correlation are marked with an `x'.}
\label{align}
\end{figure}


\begin{table}[]
\caption{Results for Ablation Study.}
\begin{center}
\begin{tabular}{lcccc}
\toprule
                                  & \multicolumn{2}{c}{\textbf{EER(\%)}} & \multicolumn{2}{c}{\textbf{min t-DCF}} \\ \cmidrule{2-5}
\multirow{-2}{*}{\textbf{Activation function}} & avg               & best             & avg                & best              \\ \midrule
ReLU                           & 1.18              & \textbf{1.13}             & 0.0388             & 0.0362            \\ \cmidrule{1-5}
Exponential                         & \textbf{1.16}              & \textbf{1.13}             & 0.0359             & 0.0347            \\ \cmidrule{1-5}
Softplus                            & 1.21              & 1.18             & \textbf{0.0340}             & \textbf{0.0276}            \\ \bottomrule
\end{tabular}
\end{center}
\label{ablation}
\end{table}

\subsection{Uncertainty Estimation}
To validate the uncertainty estimation, we plotted the results from AASIST-FADEL, showing the average uncertainty on the x-axis and the EER on the y-axis across various spoofing algorithms~(A07-A19).
As illustrated in Fig. \ref{align}, there is a strong correlation between the average uncertainty and EER, except for the cases of A07, A16, and A17.
This correlation demonstrates that the uncertainty effectively reflects the model's confidence in its predictions.

\subsection{Ablation Study}
Our proposed method ensures the non-negativity of evidence by utilizing an activation function.
We conducted an ablation study to investigate the impact of different activation functions.
Table \ref{ablation} presents the EER and min t-DCF results for each activation function applied to AASIST-FADEL.
For EER, the exponential function yielded the best performance, while for min t-DCF, the softplus function outperformed the others.

\section{Conclusion}
We proposed FADEL, an uncertainty-aware fake audio detection framework using evidential deep learning.
By integrating the model uncertainty into predictions, FADEL could mitigate the overconfidence problem and generate more reliable predictions in OOD spoofing attacks.
Experimental results demonstrate that FADEL significantly improved the performance of baseline models.

\bibliography{refs}
\bibliographystyle{IEEEtran}


\end{document}